\documentclass[]{spie}  

\usepackage{amsmath,amsfonts,amssymb}
\usepackage{graphicx}
\usepackage[colorlinks=true, allcolors=blue]{hyperref}
\usepackage{placeins}
\usepackage{subcaption}

\title{Exploration of small footprint stick-slip piezoelectric actuators for use in the FLEX fibre positioner system for the Wide-field Spectroscopic Telescope (WST)}

\author[a]{Joseph W. Barrow}
\author[a]{Suryansh Saxena}
\author[a]{Jon S. Lawrence}
\author[b]{Aaron Omadutt}
\author[b]{Roelof S. de Jong}
\affil[a]{Australian Astronomical Optics, Macquarie Univ. (Australia)}
\affil[b]{Leibniz-Institut für Astrophysik Potsdam (Germany)}

\authorinfo{Further author information: (Send correspondence to J. W. Barrow)\\J. W. Barrow: E-mail: joe.barrow@mq.edu.au}

\begin{document} 
\maketitle

\begin{abstract}
This work presents a novel stick-slip piezoelectric actuator for the FLEX fibre positioner, addressing the challenge of reducing footprint while maintaining precision in next-generation multi-object spectroscopic instruments such as the Wide-field Spectroscopic Telescope (WST). The actuator combines compact geometry (3.2 mm diameter, sub 70 mm length) with high-resolution incremental motion, low power consumption, and high reliability. Three actuators can be integrated into a sub 7 mm diameter circle, enabling a simplified, single-plane assembly. This design offers a promising solution for miniaturized, accurate fibre positioning in wide-field spectroscopic applications.
\end{abstract}


\section{INTRODUCTION}\label{sec:intro}  
Multi-Object Spectrography is an essential corner stone of modern astronomy with this seen in the continuous development of MOS instruments on 4-8m telescopes such as 4MOST\cite{4MOST}, MOONS\cite{cirasuolo2020moons}, WEAVE\cite{dalton2012weave}, PFS\cite{tamura2016prime}, and DESI\cite{aghamousa2016desi} along with MOS instruments also being developed for the next generation telescopes ELT and GMT such as MOSAIC\cite{jagourel2018mosaic} and GMACS\cite{schmidt2016optical} respectively. 

With these increasing iterations of MOS instruments we see a general trend of increasing multiplexing with the under construction MUST facility having a multiplex of 20,000\cite{zhang2023conceptual}. However the proposed Wide-field Spectroscopic Telescope (WST) looks to further push that limit with a proposed 32,000 fibres feeding it's ambitious MOS capabilities\cite{bacon2024wst}. With this increasing in multiplexing, we see the fibre positioners requiring an ever decreasing pitch, putting an increasing constraint on the footprint of these fibre positioners. One of the fibre positioners being proposed for WST is the Fibre Location EXtender (FLEX) fibre positioner being developed by Leibniz-Institut f\"{u}r Astrophysik Potsdam and Australian Astronomical Optics (AAO)\cite{de2024flex} with this fibre positioner being designed for sub 7 mm pitch. Whilst this allows for the correct level of multiplexing, the requirement for three actuators for operation necessitates the development of novel small footprint actuators to allow for a single plane construction. Alongside the footprint constraints, there are several other requirements that arise for a viable actuator for the FLEX fibre positioner, with these listed below.
\begin{itemize}
    \item \textbf{Step Resolution}: 0.3 $\mu$m
    \item \textbf{Driving Force}: 1.7 N
    \item  \textbf{Maximum Stroke}: 3 mm
    \item \textbf{Velocity}: 0.2 mms$^{-1}$
    \item \textbf{Operating Frequency}: 570 Hz
    \item \textbf{Maximum Voltage}: $\leq$ 100 V
\end{itemize}
The step resolution, driving force and maximum stroke requirements arise from the operation of the FLEX positioner, whilst the velocity and operating frequency requirements arise from the WST operational targets. The step resolution requirement arises from the actuator displacement  needed to achieve positioning accuracy of $\leq$ 15 $\mu$m for $\geq$ 98\% of all active fibres. The driving force requirement arises from force required to achieve the maximum patrol radius of the FLEX when motion is such that only one of the three actuators is driving in a given direction. The maximum stroke requirement arises from the total actuator travel to achieve the full patrol radius of the FLEX and the additional defocus correction. With WST having a target repositioning time of 30 s, we can find the minimum velocity requirement of the actuator by assuming a worst case scenario where the positioner must go from one maximum patrol radius position to another, through the home position requiring a full 6 mm of travel over the repositioning time. The minimum operating frequency then arises from the frequency required when the minimum operating velocity is achieved and the assumption that only the minimum step size can be achieved. When considering the commercially available off the shelf linear actuators, we were unable to find one that met the requirements for operating the FLEX fibre positioner hence the development of custom stick-slip piezoelectric actuator shown in this study. The actuator discussed in this work has been designed such to achieve the requirements outlined including the low voltage operation with this requirement existing so as to mitigate the additional volume required for high voltage electronics. 

\section{A Summary of stick-slip actuators}\label{sec:summary}
\begin{figure}[ht]
    \centering
    \includegraphics[width=0.5\linewidth]{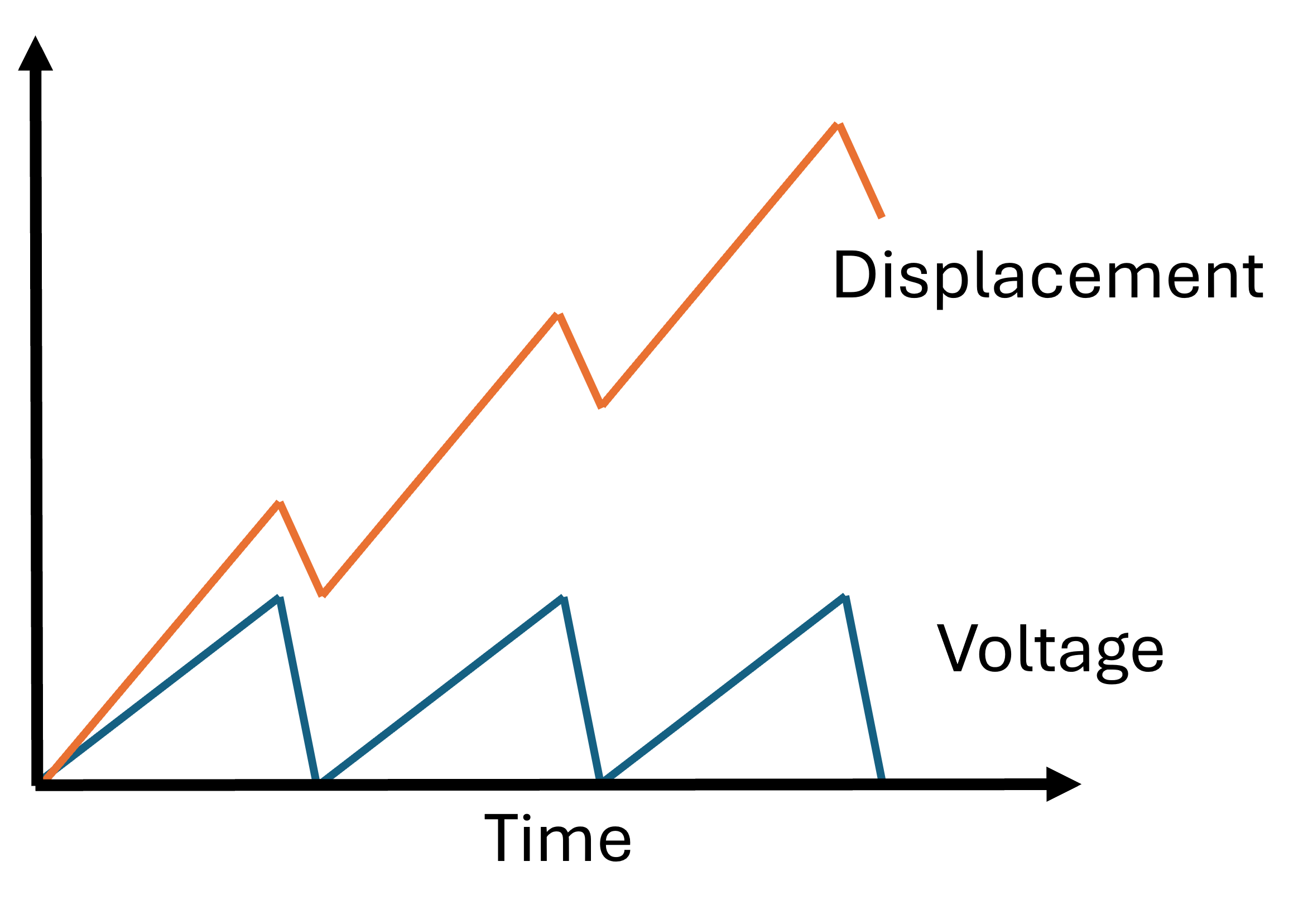}
    \caption{A simple plot of voltage against time, showing the displacement of the actuator with time also.}
    \label{fig:StickSlipAction}
\end{figure}

All piezoelectric stick-slip actuators are based around a simple principle of applying a potential difference across to a piezoelectric material, causing either a change in the length of a dimension or the material to shear. This change is then coupled with a mechanical system to either feed through a shaft or move a mass. For a stick-slip actuator, a net step is achieved through the piezoelectric motion displacing the shaft or mass during a slow "stick" phase before the piezoelectric is rapidly returned to its rest state inducing a "slip" phase. This slip phase leads to only a small inverse displacement by the mass or shaft such that the motion in the stick phase was larger than the slip phase. This process, including the voltage waveform applied to the piezoelectric actuator, can be seen in Figure \ref{fig:StickSlipAction}. For systems that are held in a fixed position for prolonged periods of time, a stick-slip actuator is therefore a considered choice due to the actuator being able to hold positions in a no-power condition.

\begin{figure}[h]
    \centering
    \includegraphics[width=\linewidth]{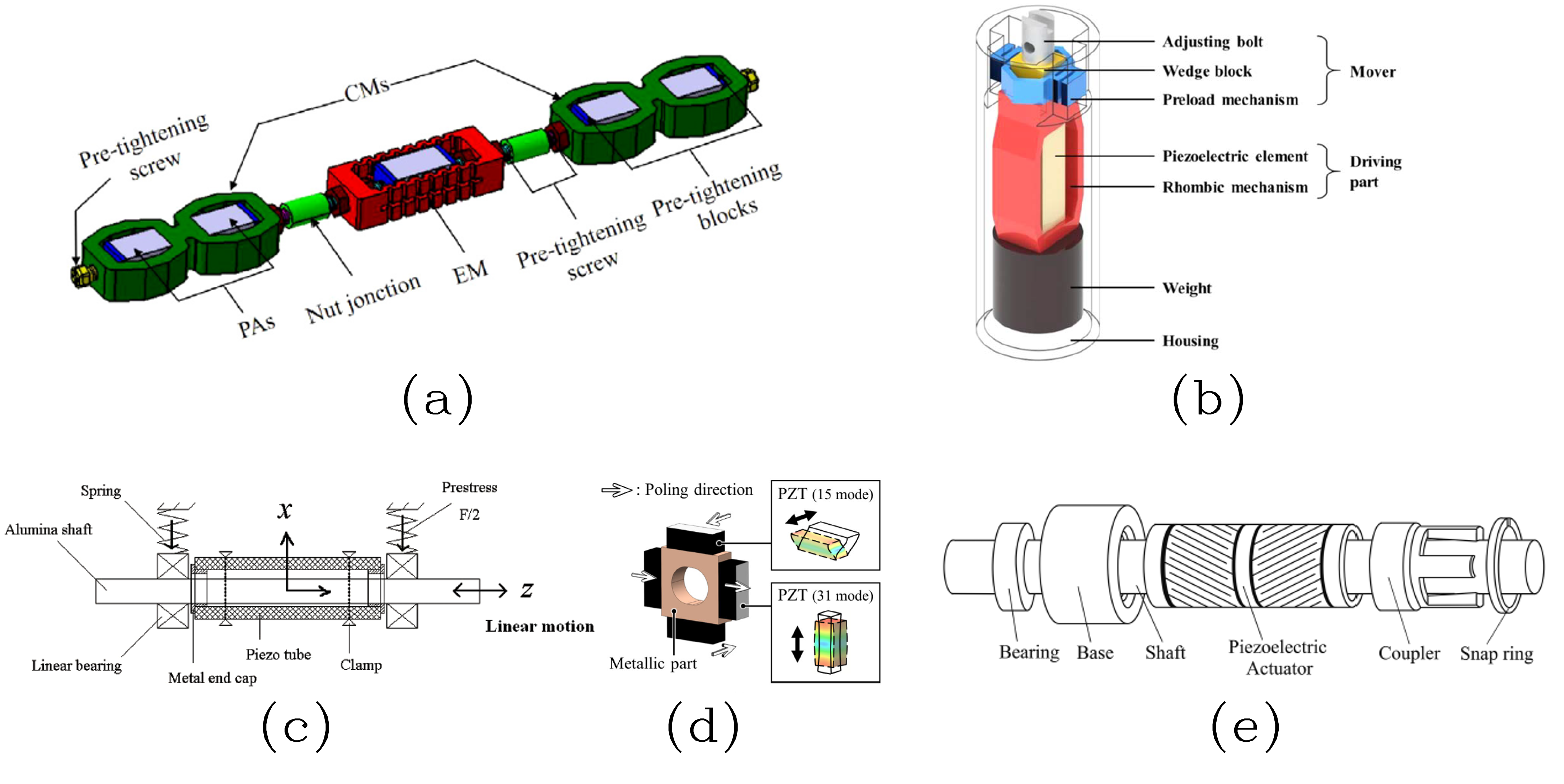}
    \caption{Examples of styles of piezoelectric actuators from the literature where (a) shows the inchworm design of Genna et al \cite{inchworm}, (b) the inertial impact actuator developed by Shao et al \cite{impact}, (c) and (d) the linear bending modes actuators from Guo et al \cite{guo2014_bending} and Izuhara et al \cite{Izuhara_bending} respectively, and (e) the two-degree of freedom torsional linear actuator developed by Han et al \cite{2dof}.}
    \label{fig:TradeStudyActuators}
\end{figure}

When developing a linear piezoelectric stick-slip actuator, there are several different approaches taken within the literature with each of these being considered for this actuators design. The most common of these approaches are the inchworm, inertial impact, linear bending modes, and mechanical clamping. Examples from the literature of an inchworm, inertial impact, linear bending modes, and a two-degree of freedom, torsional piezoelectric actuators can be seen in Figure \ref{fig:TradeStudyActuators}. By comparing the examples in the literature against the requirements of the actuator for driving the FLEX fibre positioner, a design philosophy was developed.

Whilst the actuator velocity required for this work was well below the maximum for many of the designs seen in the literature, it was the factor that rules out the inchworm style designs with speeds of the order $\mu$ms$^{-1}$ not uncommon \cite{inchworm} due to three separate elements needing to be actuated during motion. It should be noted that for applications where a large weight must be supported with a low operating velocity, a inchworm design is ideal due to the large blocking force achieved through two clamping segments. When considering actuators that take advantage of linear bending modes of a piezoelectric, such as the works of Guo et al\cite{guo2014_bending} and Izuhara et al\cite{Izuhara_bending}, we see that these devices can often be very compact and offer very high velocities. This led to their use in astronomical instrumentation\cite{Younes} through Squiggle motors, developed by NewScale Technologies, which also offer sub-micron resolution. However, this family of linear piezoelectric stick-slip actuators struggle to achieve the required drive force for this application within the tight footprint that is required. When exploring concepts such as the inertial impact designs in the work of Shao et al \cite{impact}, we see that due to this type of actuator having a mechanical amplifier frame they are not suitable for this project as these amplifier elements lead to a footprint that is too large for this application. The final design principle that was considered but ultimately not chosen was the torsional actuator from Han et al \cite{2dof}. This actuator used novel dual helical electrodes to allow for a linear actuation with no rotational elements however, due to the complexity of the design, the voltages required to achieve the required drives forces and step sizes were large whilst the pair of helical electrodes adds additional levels of electrical complexity. Therefore, this then leaves a mechanical clamping actuator design. 

\section{Design of a piezoelectric actuator system}\label{sec:design}
When approaching the design of a piezoelectric actuator for the FLEX positioner and the requirements for WST, there are three key design areas. In this section each of these will be addressed from the mechanical design of the preloading mechanism, the choice of the piezoelectric element, and the combining of these actuators to operate the FLEX positioner. 

\subsection{Mechanical Preload Design}\label{sec:preload}
The mechanical clamping design used in this work uses the deflection of an element to make contact with the drive shaft. Through designing the geometry of the deflecting elements, the preload is then the force required to make contact with the shaft. When designing a stick-slip linear actuator the system must have a load applied to it, a preload, with the magnitude of this being dependant on both the target drive force of the actuator and the friction coefficient between the shaft and preload mechanism. For an actuator with a cylindrical shaft, the preload would be applied radially and for a drive force tangential to this. The drive force is then given by
\begin{equation}
    F_{d} = \mu F_{p},
\end{equation}
where $F_{d}$ is the drive force, $F_{p}$ the preloading force and $\mu$ the friction coefficient. The preload in this design is applied by several individual fingers that deflect inwards in a manner similar to a collet, with this deflection being applied through an outer-sleeve that has a decreased inner diameter at the required loading point. Through careful design of the fingers' geometries, the fingers can be such that when they are deflected to make contact with the drive shaft, the force required to achieve this is the desired preload. This leads to a fixed and constant preload being applied to the fingers. We would therefore expect a consistent drive force to be maintained through the lifetime of the actuator as only surface wear on the fingers or shaft would be expected to alter the preload, and with a high tensile stress, high wear resistance material with a good fatigue life, we would expect this to be negligible. For the actuator to be considered as stick-slip, it must have a low friction coefficient in order to allow the slip phase, with a $\mu\leq0.1$ to allow for fine positioning. The actuator has been designed such that no external lubrication is required, meaning that a high surface finish combined with a low friction coating must be used to achieve the stick-slip friction contact.

\begin{figure}[h]
    \centering
    \includegraphics[width=0.4\linewidth]{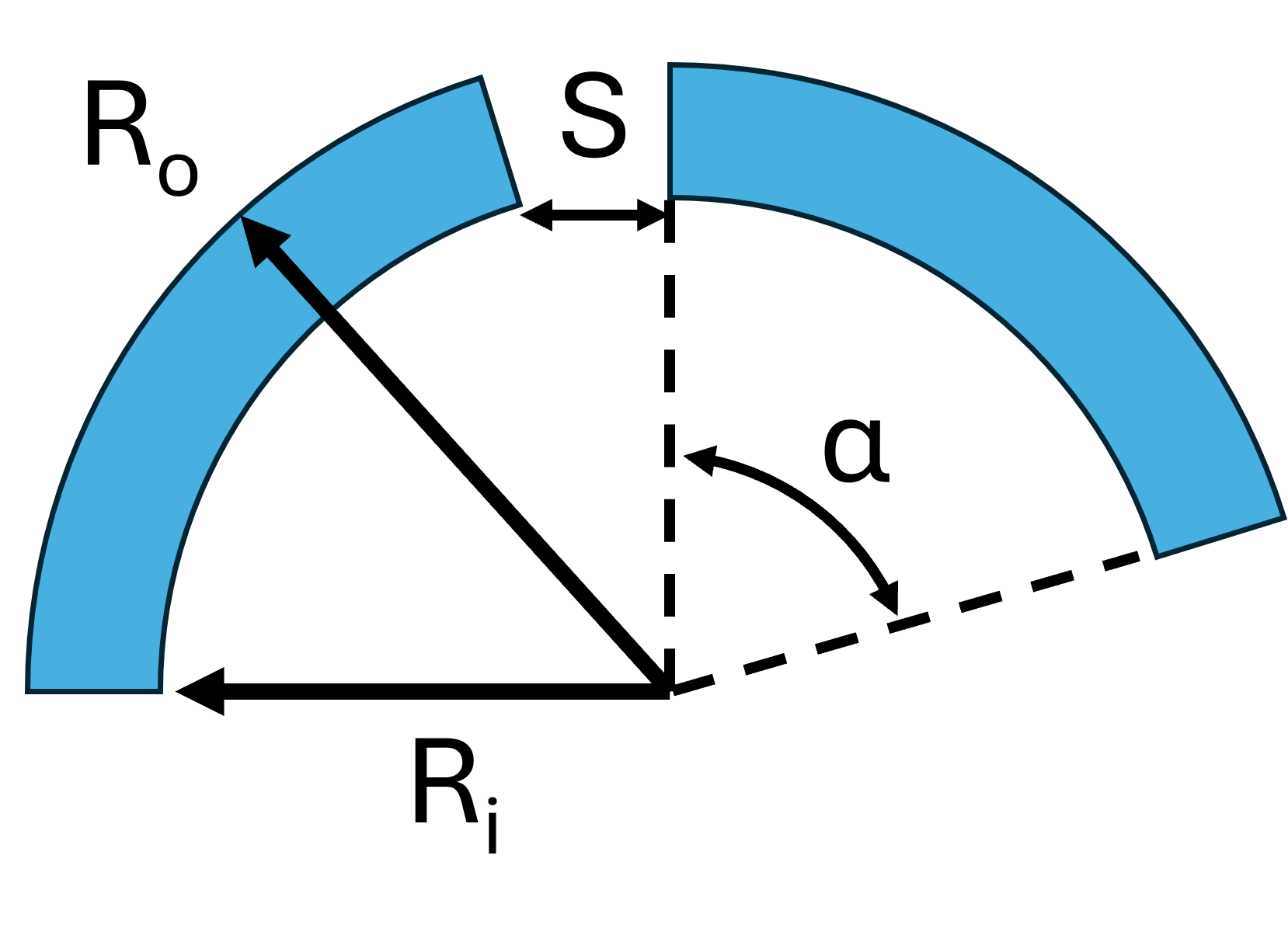}
    \caption{A diagram of the cross section of a finger of the preload mechanism, outlining the inner and outer radii, the slot gap between fingers and angle segment each finger occupies.}
    \label{fig:Finger_definition}
\end{figure}

In order to reduce the stress and force requirement per finger, the preloader is designed to have multiple fingers, with the cross section of each element then being a segment of an annulus having an inner and outer radius, with a cross section shown in Figure \ref{fig:Finger_definition}. 

\begin{figure}[h]
    \centering
    \includegraphics[width=0.75\linewidth]{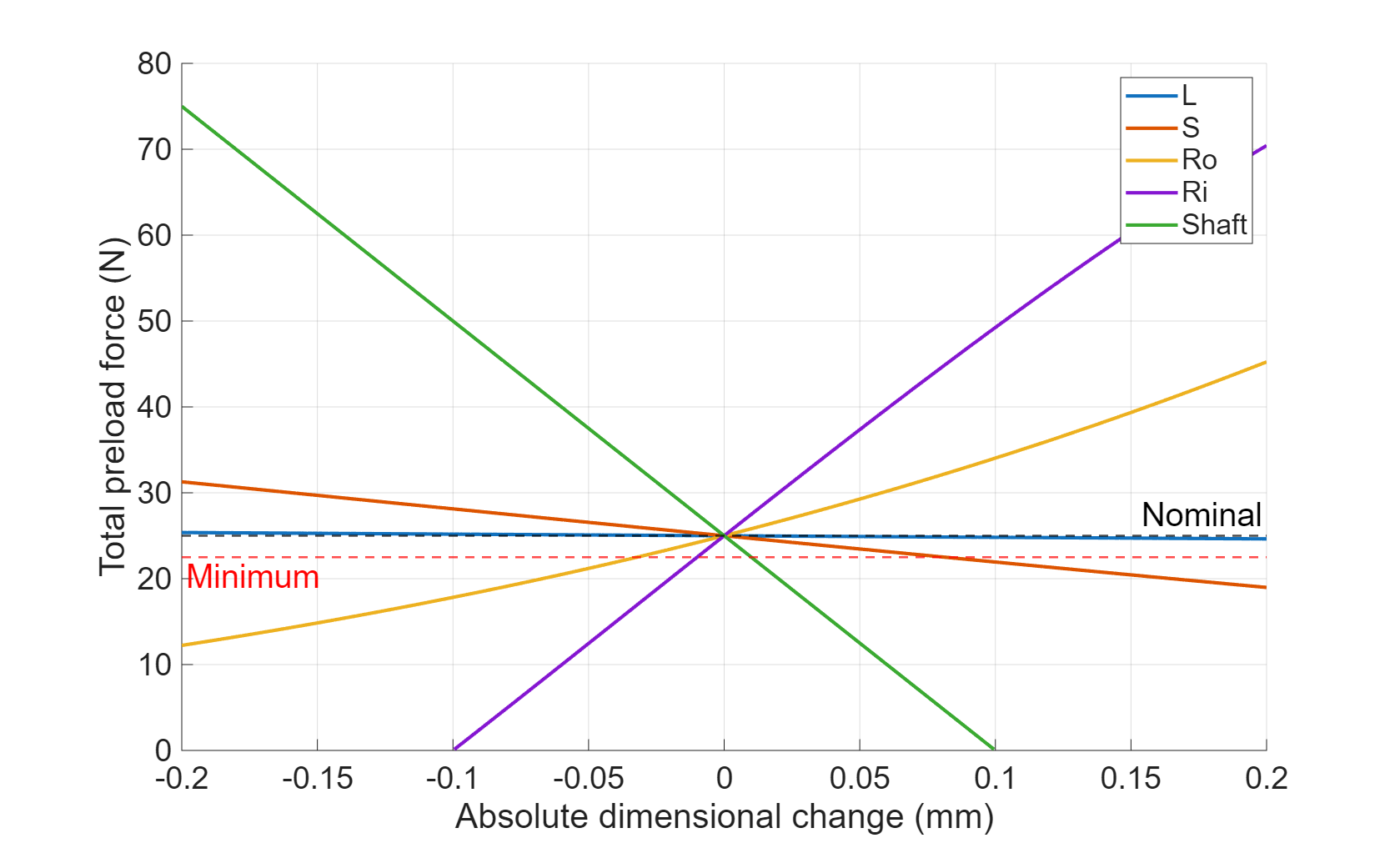}
    \caption{A comparison of impact on the preload force across the key dimensions, when only one is varied.}
    \label{fig:Dimension_Impacts}
\end{figure}

Each finger's length is then determined using Euler-Bernoulli beam theory from the following expression in which $\delta$ is the deflection of the finger, $E$ is the Young’s Modulus of the material, $F_{n}$ is the preload force required per $n$ fingers and $I$ is the second moment of area which is a function of the finger’s cross section, in which $a$ is the fractional length at which the finger pinches the shaft
\begin{equation}
    L = \frac{1}{a}\sqrt[3]{\frac{3EI\delta}{F}}.
\end{equation}
Using this expression we can define the geometry of the preload mechanism whilst also exploring the impact of each dimension on the preload force, outlining the key geometries for the machining of the preload mechanism. From Figure \ref{fig:Dimension_Impacts} we can see that for a target geometry, the most crucial dimensions are the inner radius of the finger $R_{i}$ and radius of the drive shaft. This is to be expected, as these directly impact the deflection of the fingers, with the inner radius also determining the wall thickness and hence second area moment of the finger. These dimensions will therefore need the tightest tolerance during machining with sub 10 $\mu$m being required, whilst for dimensions such as the length, a much looser tolerance of the order 0.2 mm can be used. Through this analysis, we can evaluate the lower sensitivity geometries and allow for lower cost machining tolerances where possible.

\subsection{Piezoelectric Design}\label{sec:PiezoChoice}
A key aspect of a stick/slip piezoelectric actuator is the piezoelectric element which is used to move the shaft and hence the preloader during the stick section of the motion. The first decision for the piezoelectric element was whether to go for a stack or a tube. For this actuator, a piezoelectric stack was chosen over a piezoelectric tube. The reason for this was the commercial availability of stacks with small footprints which was not the case for tubes. Whilst a custom tube could be included in a future development of the actuator, the option for an off-the-shelf piezoelectric stack which could be verified against a data-sheet for reliable performance was the preferable option. 

When considering a piezoelectric there are four main parameters that need to be considered: the footprint, the maximum stroke, the maximum voltage, and the capacitance. Firstly the footprint must be one that meets the space envelope requirements for the actuator. The maximum stroke dictates how much the piezoelectric can extend when a given voltage is applied, whilst the maximum voltage states the maximum voltage that the piezoelectric is rated to operate at. Supplying a voltage less than this would likely increase the lifespan of the piezoelectric but at the cost of the available stroke. Finally the, capacitance plays an important role in the maximum operating frequency of the piezoelectric. This is due to the relation between the current $I$ and voltage $V$ supplied to the piezoelectric, the frequency $f$ this signal is applied at, and the capacitance $C$ of the stack. These are related by
\begin{equation}
    I = 2\pi f C V
\end{equation}

\begin{figure}
 \begin{subfigure}{0.49\textwidth}
     \includegraphics[width=\textwidth]{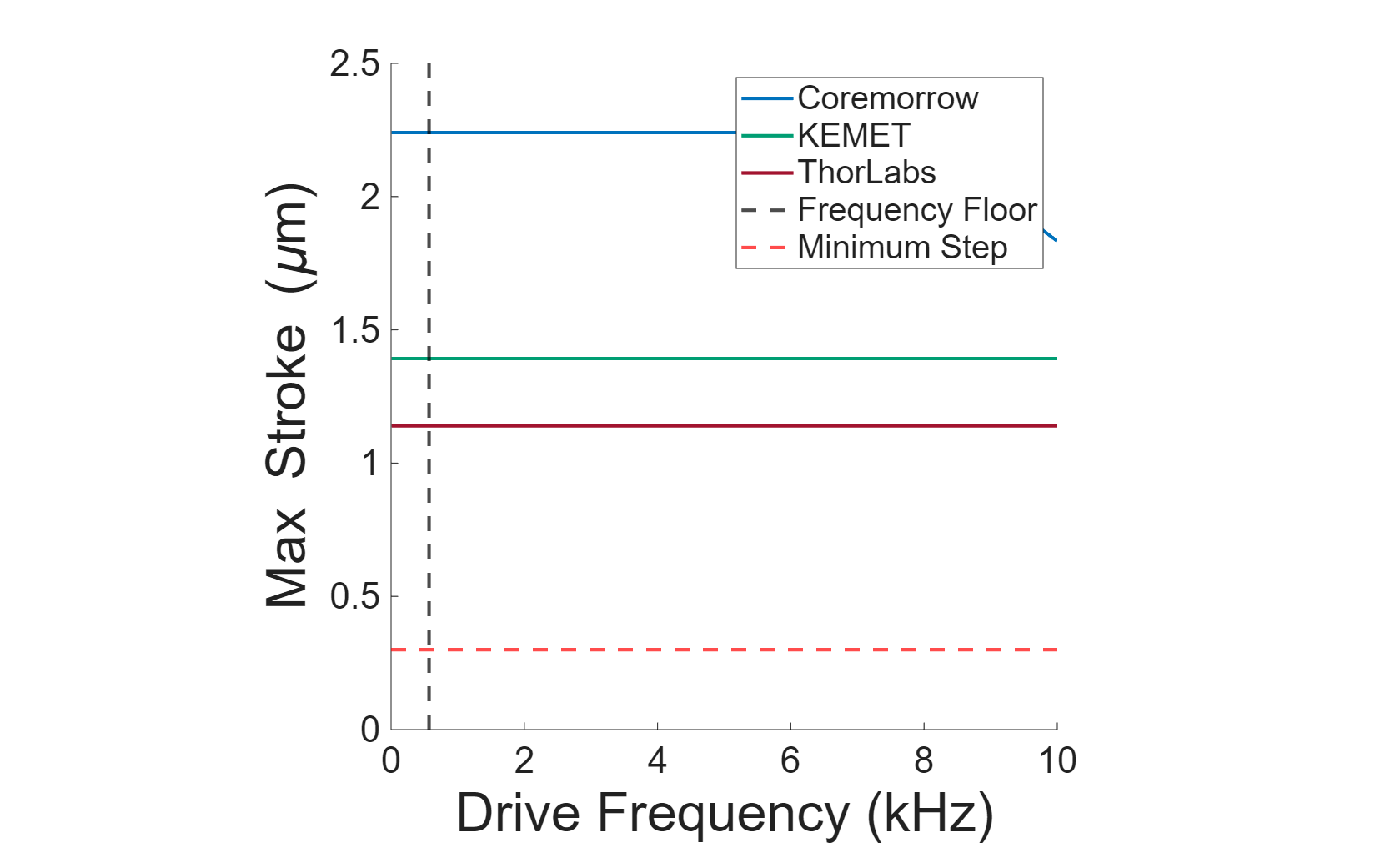}
     \caption{12 V}
     \label{fig:a}
 \end{subfigure}
 \hfill
 \begin{subfigure}{0.49\textwidth}
     \includegraphics[width=\textwidth]{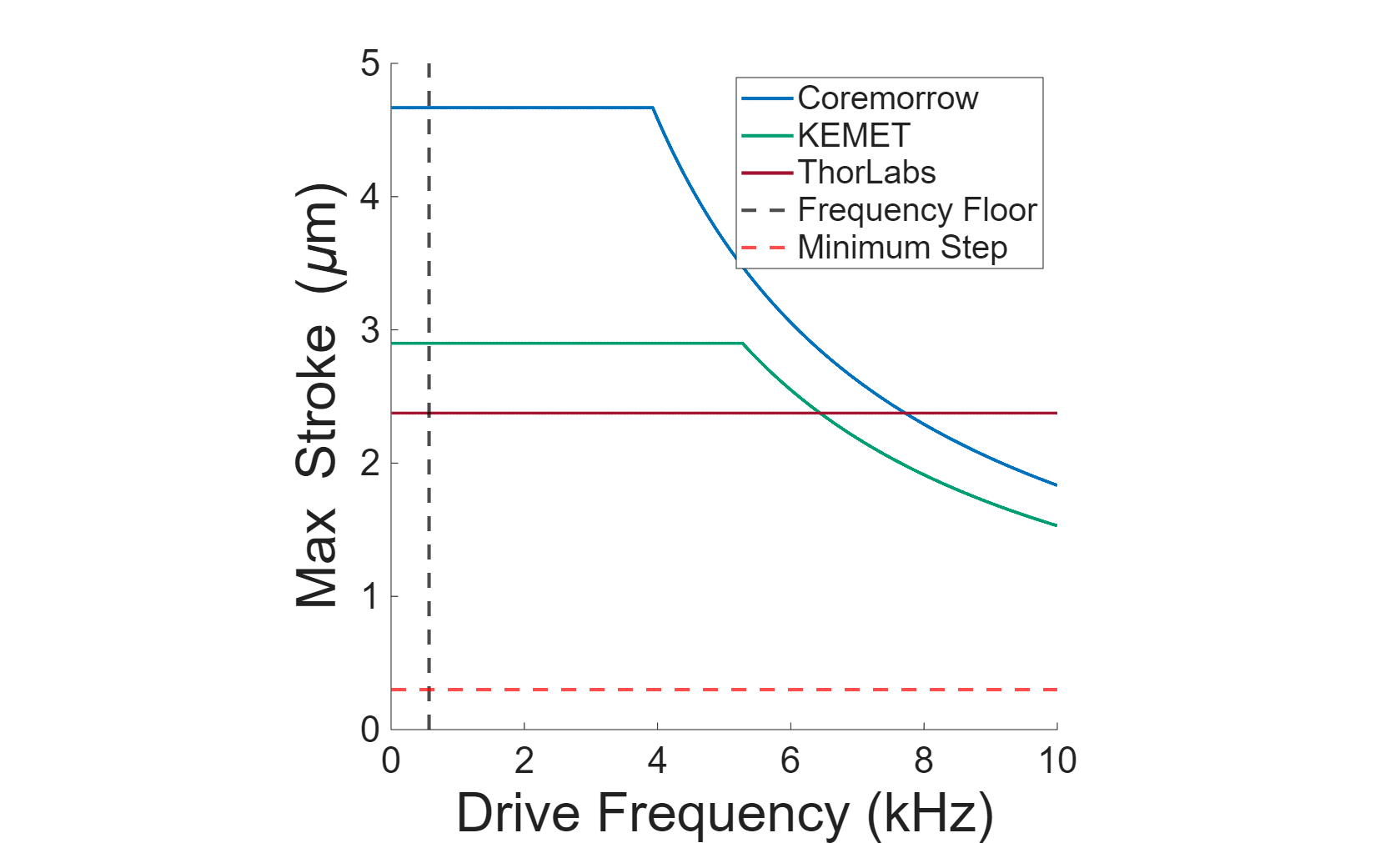}
     \caption{25 V}
     \label{fig:b}
 \end{subfigure}
 
 \medskip
 \begin{subfigure}{0.49\textwidth}
     \includegraphics[width=\textwidth]{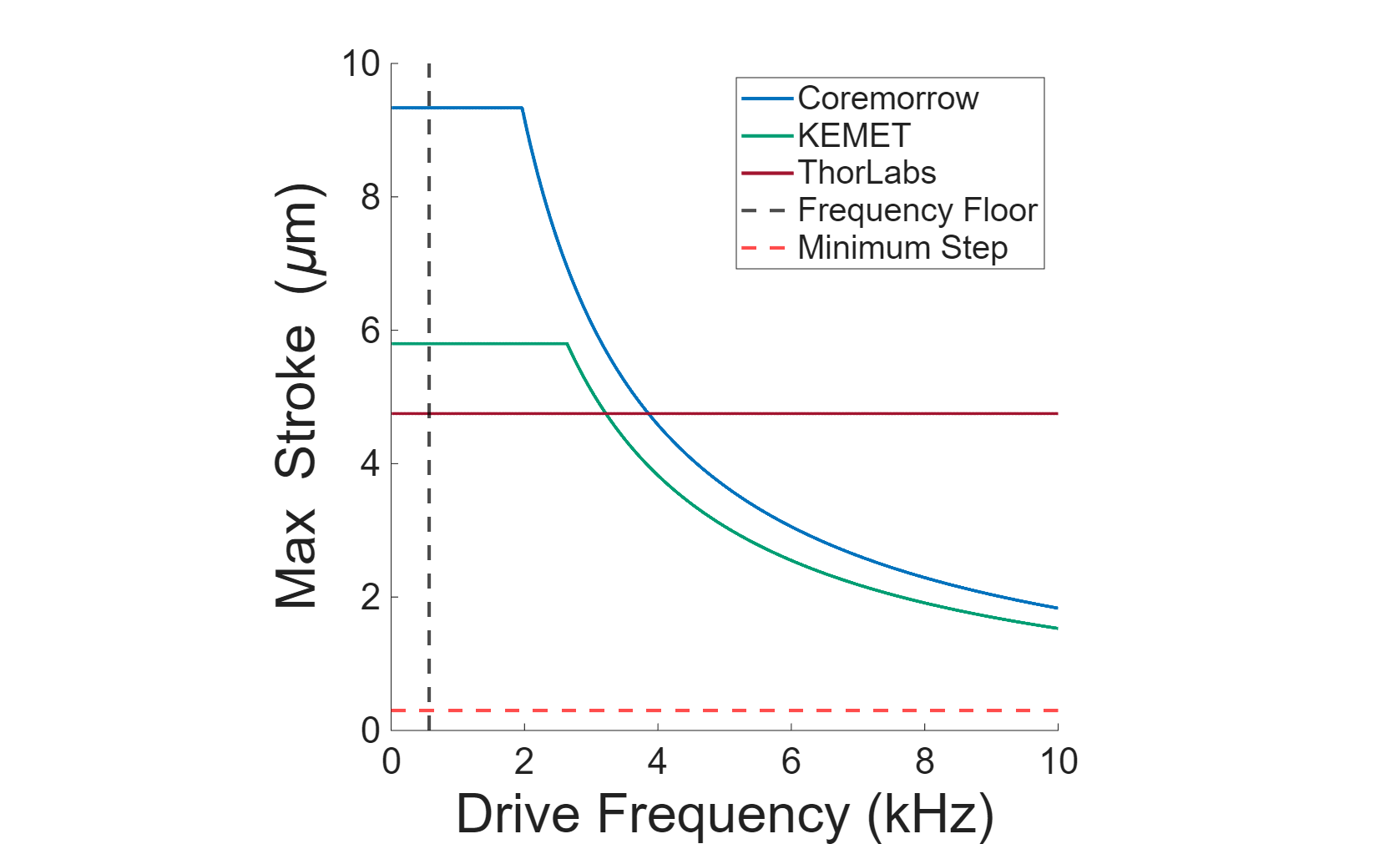}
     \caption{50 V}
     \label{fig:c}
 \end{subfigure}
 \hfill
 \begin{subfigure}{0.49\textwidth}
     \includegraphics[width=\textwidth]{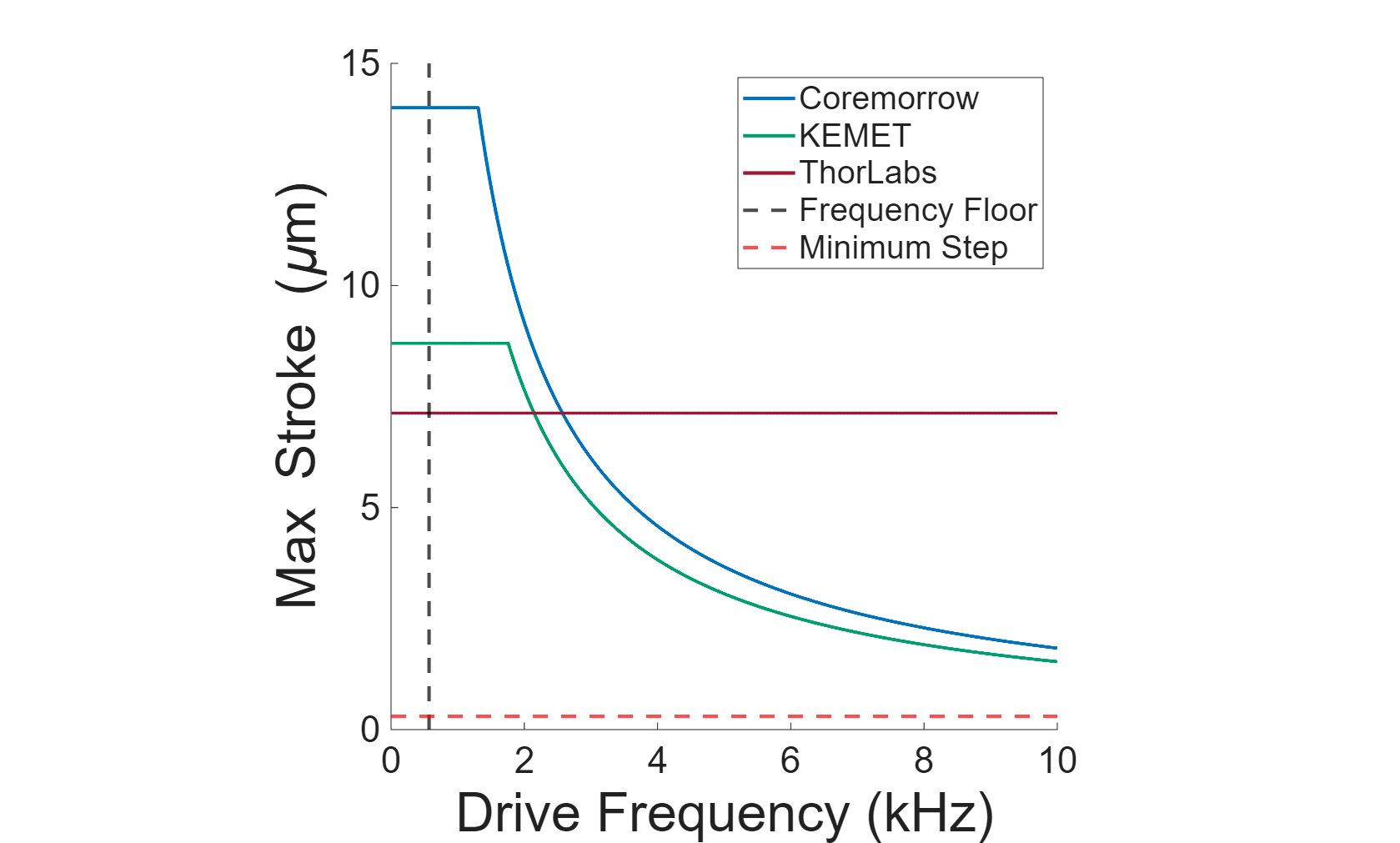}
     \caption{75 V}
     \label{fig:d}
 \end{subfigure}
 \caption{Maximum stroke for a range of piezoelectric stacks when driven at a range of different voltages with these being plotted as functions of frequency. The caption for each sub figure indicates the applied voltage. The maximum stroke is taken to be a linear interpolation of the stroke based on the maximum stroke at the maximum operating voltage.}
 \label{fig:Piezo_Comparisons}
\end{figure}

The maximum available current from the chosen amplifier used during testing is 280 mA and in order to achieve the repositioning goals of the focal-surface, the minimum operating frequency must be 400 Hz. Using these parameters, we compared three different piezoelectric stacks: the PSt150/2x3/20H from Coremorrow, the AE0203D16 from KEMET-TOKIN, and the PK3CMP1 from Thorlabs. For a given voltage, the maximum extension the stack can achieve is then plotted as a function of frequency, where the step will decay as the maximum voltage decreases with increasing frequency. From this we can determine the usable frequency range of the stacks. A comparison of the three stacks across a range of voltages can be seen in Figure \ref{fig:Piezo_Comparisons}. From comparing the performance of the stacks across a range of voltages and frequencies, the Coremorrow piezoelectric stack (PSt150/2x3/20H) will be used as the baseline designs, due to the largest stroke across the majority of the frequency range, allowing for a higher frequency operation should this be required, whilst still offering larger stroke as the voltage decreases. We can also see that due to the low frequency and step resolution requirements, there are a large range of piezo stacks that would be available and further miniaturization could be possible. For a high frequency control scheme where a larger minimum stroke is required, the ThorLabs PK3CMP1 piezoelectric stack would be a preferred candidate due to the lower capacitance and hence wider operating frequency showing a flat-band response through 10 kHz. 

\subsection{An Actuator System for the FLEX Positioner}\label{sec:congiguration}

\begin{figure}[h]
    \centering
    \includegraphics[width=\linewidth]{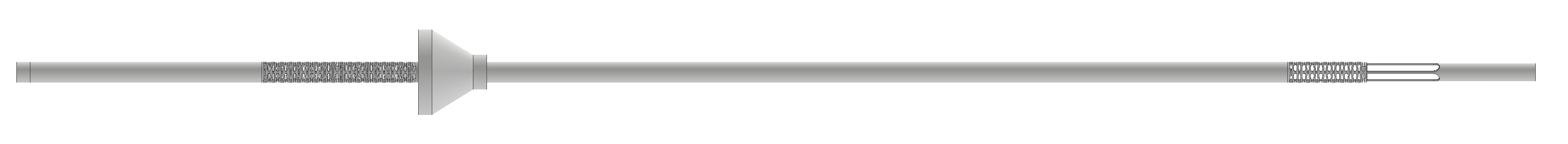}
    \caption{A diagram of the FLEX positioner \cite{FLEX_Aaron}.}
    \label{fig:FLEX}
\end{figure}

\begin{figure}[h]
    \centering
    \includegraphics[width=0.6\linewidth]{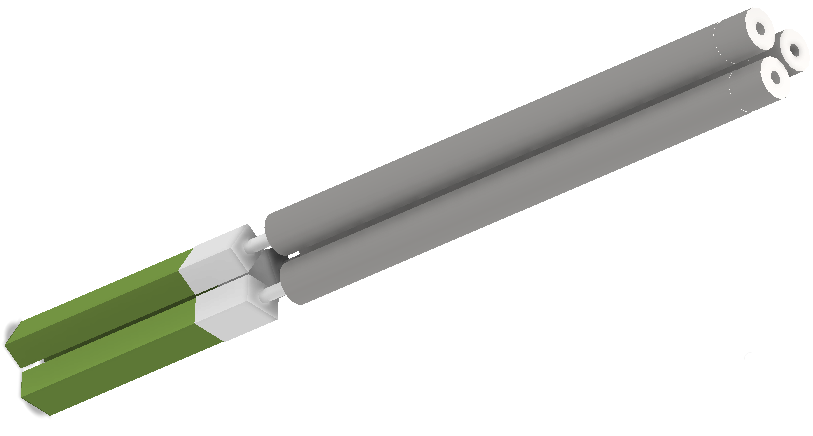}
    \caption{An outline of how three actuators may be configured so as to maintain the pitch and tiling of FLEX positioner for WST.}
    \label{fig:ActuatorLayout}
\end{figure}

\begin{figure}[h]
    \centering
    \begin{subfigure}{0.9\textwidth}
    \includegraphics[width=\linewidth]{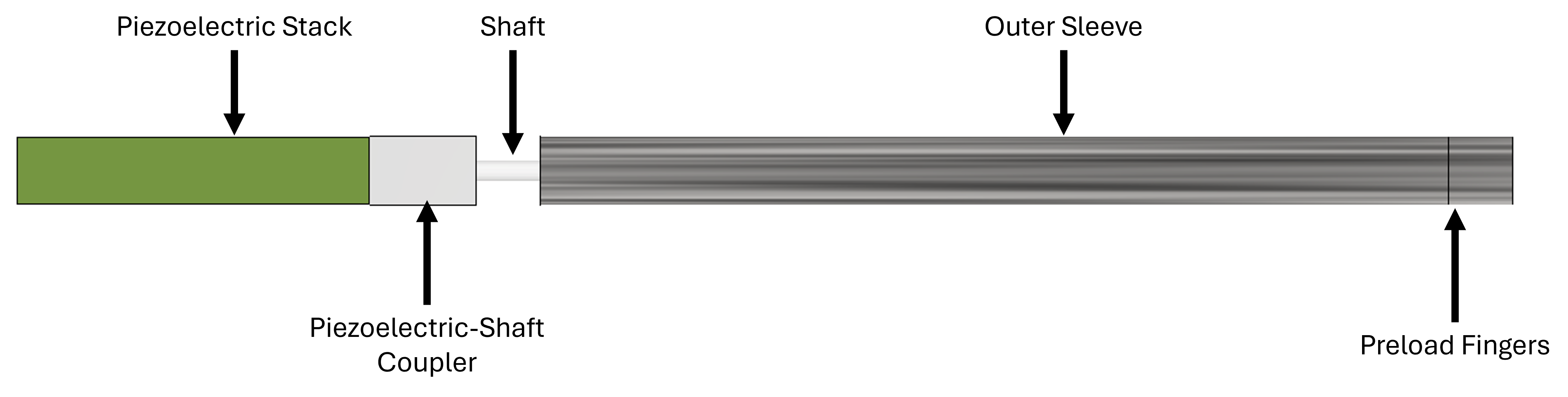}
    \caption{}
    \label{fig:SingleActuator}
    \end{subfigure}
    \hfill
    \begin{subfigure}{0.9\textwidth}
    \includegraphics[width=\linewidth]{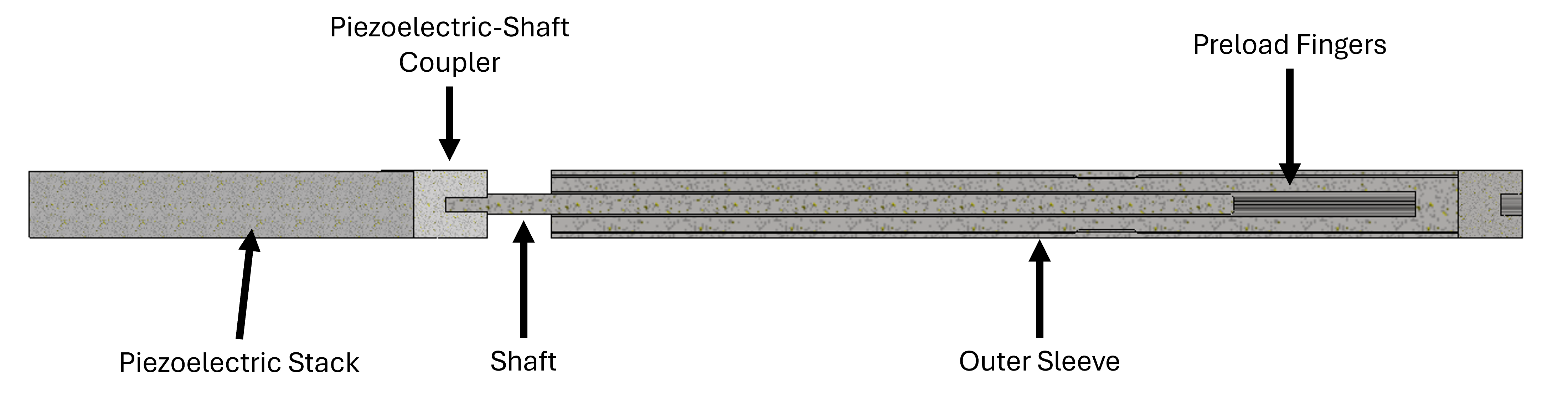}
    \caption{}
    \label{fig:SingleActuator Crossssection}    
    \end{subfigure}
    \hfill
    \caption{Diagrams of the schematics of the actuator developed for the FLEX positioner, with (a) showing a labelled diagram of the constructed positioner whilst (b) shows the cross section.}
\end{figure}

For the FLEX positioner as seen in Figure \ref{fig:FLEX}, three actuators combine to induce the X-Y patrol motion of the positioner, alongside Z-axis defocus correction. With the pitch of the positioner being sub 7 mm, the three actuators must be configured in a way so as to maintain this pitch, whilst also allowing tiling between positioners across the plane and fibre routing between the actuators. An example of how the actuators may be configured for a single FLEX positioner can be seen in Figure \ref{fig:ActuatorLayout}, with a diagram of a single actuator shown in Figure \ref{fig:SingleActuator} with Figure \ref{fig:SingleActuator Crossssection} showing the cross-section. By allowing the fibre to be routed at the mid point between the actuators, no additional strains or torsions need be applied to the fibre when transitioning from the FLEX positioner mechanism, through the actuators and to the wider fibre system. 

Through designing the actuators to be a linear actuator, this reduces some of the electrical complexity that could arise from the FLEX positioner being a multi-actuator system. With each piezo stack only being excitable in the vertical direction, only 2 electrodes and a single signal can be used to drive each actuator. When driving multiple piezoelectric actuators in unison, there is a risk for crosstalk effects across the focal plane due to coupled vibrations. Due to the tripod configuration the FLEX positioner uses, a preferential orthogonal operating grid doesn't exist. This acts to further reduce crosstalk effects as positioners are free to be driven to their target in direct paths where appropriate. A further means to reduce the coupled vibrations when multiple actuators are being driven in unison can be achieved through drive signals being used with differing phases. With a defocus element being required as the FLEX positioner patrols on the focal surface, a drive scheme for the positioner can also be considered in which no more than two actuators are required to be driven at any time, which further increases the power efficiency and ease of operation.

\section{Conclusion}
Through considering the requirements of operating the FLEX positioner for the proposed Wide-field Spectroscopic Telescope, a novel stick-slip piezoelectric actuator has been designed. Due to the small space envelope, step resolution, and moderate drive force a custom linear actuator was required due to a lack of commercially available options. A stick-slip preload design was chosen to allow for a no-power hold condition offering a low power control scheme, a crucial consideration for a three actuator positioner in a 30,000 multiplexing multi-object instrument. A mechanical preload which requires no tuning should provide a low part count, consistent drive force across the actuator lifetime. The next steps for this project are the prototyping of an over-scale version of the actuator for an increased handle-ability. This allows for an integration with FLEX positioner as a control scheme test case before a correctly scaled actuator is then prototyped and tested. 

\acknowledgements
This research was funded by the Australian Government through the Australian Research Council and their Discovery Project Scheme (DP250103698). This research was supported by the Leibniz Competition grant K593/2024.  This project has received funding from the European Union Horizon Europe Research and Innovation Action under grant agreement no. 101183153 -WST. Views and opinions expressed are however those of the author(s) only and do not necessarily reflect those of the European Union or the European Research Executive Agency (REA). Neither the European Union nor the REA can be held responsible for them.

\bibliography{report} 
\bibliographystyle{spiebib} 

\end{document}